\documentstyle[epsf]{mn}
\newif\ifAMStwofonts
\AMStwofontstrue

\ifoldfss
  \ifCUPmtlplainloaded \else
    \NewTextAlphabet{textbfit} {cmbxti10} {}
    \NewTextAlphabet{textbfss} {cmssbx10} {}
    \NewMathAlphabet{mathbfit} {cmbxti10} {} 
    \NewMathAlphabet{mathbfss} {cmssbx10} {} 
  \fi
  \ifAMStwofonts
    \ifCUPmtlplainloaded \else
      \NewSymbolFont{upmath} {eurm10}
      \NewSymbolFont{AMSa} {msam10}
      \NewMathSymbol{\upi}     {0}{upmath}{19}
      \NewMathSymbol{\umu}     {0}{upmath}{16}
      \NewMathSymbol{\upartial}{0}{upmath}{40}
      \NewMathSymbol{\leqslant}{3}{AMSa}{36}
      \NewMathSymbol{\geqslant}{3}{AMSa}{3E}

       \let\le=\leqslant
       
    \fi
  \fi
\fi 

\ifnfssone
  \newmathalphabet{\mathit}
  \addtoversion{normal}{\mathit}{cmr}{m}{it}
  \addtoversion{bold}{\mathit}{cmr}{bx}{it}
  \newmathalphabet{\mathbfit} 
  \addtoversion{normal}{\mathbfit}{cmr}{bx}{it}
  \addtoversion{bold}{\mathbfit}{cmr}{bx}{it}
  \newmathalphabet{\mathbfss} 
  \addtoversion{normal}{\mathbfss}{cmss}{bx}{n}
  \addtoversion{bold}{\mathbfss}{cmss}{bx}{n}
  \ifAMStwofonts
    \ifCUPmtlplainloaded \else
      \UseAMStwoboldmath
      \makeatletter
      \new@mathgroup\upmath@group
      \define@mathgroup\mv@normal\upmath@group{eur}{m}{n}
      \define@mathgroup\mv@bold\upmath@group{eur}{b}{n}
      \edef\UPM{\hexnumber\upmath@group}
      \new@mathgroup\amsa@group
      \define@mathgroup\mv@normal\amsa@group{msa}{m}{n}
      \define@mathgroup\mv@bold\amsa@group{msa}{m}{n}
      \edef\AMSa{\hexnumber\amsa@group}
      \makeatother
      \mathchardef\upi="0\UPM19
      \mathchardef\umu="0\UPM16
      \mathchardef\upartial="0\UPM40
      \mathchardef\leqslant="3\AMSa36
      \mathchardef\geqslant="3\AMSa3E

       \let\le=\leqslant

    \fi
  \fi
\fi 

\ifnfsstwo
  \DeclareMathAlphabet{\mathbfit}{OT1}{cmr}{bx}{it}
  \SetMathAlphabet\mathbfit{bold}{OT1}{cmr}{bx}{it}
  \DeclareMathAlphabet{\mathbfss}{OT1}{cmss}{bx}{n}
  \SetMathAlphabet\mathbfss{bold}{OT1}{cmss}{bx}{n}
  \ifAMStwofonts
    \ifCUPmtlplainloaded \else
      \DeclareSymbolFont{UPM}{U}{eur}{m}{n}
      \SetSymbolFont{UPM}{bold}{U}{eur}{b}{n}
      \DeclareSymbolFont{AMSa}{U}{msa}{m}{n}
      \DeclareMathSymbol{\upi}{0}{UPM}{"19}
      \DeclareMathSymbol{\umu}{0}{UPM}{"16}
      \DeclareMathSymbol{\upartial}{0}{UPM}{"40}
      \DeclareMathSymbol{\leqslant}{3}{AMSa}{"36}
      \DeclareMathSymbol{\geqslant}{3}{AMSa}{"3E}

       \let\le=\leqslant

    \fi
  \fi
\fi 

\ifCUPmtlplainloaded \else
  \ifAMStwofonts \else 
    \def\upi{\pi}
    \def\umu{\mu}
    \def\upartial{\partial}
  \fi
\fi

\title{Segregated Optical$-$NIR colour distributions of MDS galaxies}
\author[] {I. Ferreras$^{1,2}$, L. Cay\'on$^2$, 
	E. Mart\'\i nez-Gonz\'alez$^2$,	N. Ben\'\i tez$^{3}$ \\
1. Departamento de F\'\i sica Moderna, Universidad de Cantabria, 
	39005 Santander, Spain.\\
2. Instituto de F\'\i sica de Cantabria, Fac. Ciencias, Av. los
	Castros s/n, 39005 Santander, Spain\\
3. Astronomy Department, University of California, Berkeley, 
	CA 94720, USA.\\}
\date{\today}

\pagerange{\pageref{firstpage}--\pageref{lastpage}}
\pubyear{1998}

\begin{document}

\maketitle

\label{firstpage}

\begin{abstract}
We present a $K$ survey of 29 fields covering approximately $90$ arcmin$^2$ 
from the Medium Deep Survey (MDS) catalogue down to a completeness
magnitude of $K=18.0$ (limiting magnitude $K=19.0$). The 
morphology obtained by the MDS team using high resolution images from 
HST/WFPC2 along with our NIR observations allow a Colour--Magnitude and 
Colour--Colour analysis that agrees in general with spectral evolution 
models \cite{bruz98} especially if a reasonable range of 
metallicities for the Simple Stellar Populations used 
($0.2 < Z/Z_\odot < 2.5$) is considered. However, a significant 
population of {\it spheroids} was found, which appear {\it bluer than 
expected}, confirming previous observations (Forbes et al. 1996, 
Koo et al. 1996, Glazebrook et al. 1998). This blueness might possibly 
signal the
existence of non-negligible star formation in ellipticals and bulges
at medium redshift. A number counts calculation for different morphological
types show disks become the dominant population at faint magnitudes.
The median redshift of the sample is $z\sim 0.2$, from a
photometric redshift estimation using $V-K$ and $I-K$. A search for 
EROs in the survey field was also performed, with no detection of objects 
having $I-K>4.5$, setting an upper limit to the number density of EROs at
$dn_{\rm EROs}/d\Omega < 0.011$ arcmin$^{-2}$ ($K\le 18.0$). 
\end{abstract}


\begin{table}
  \begin{center}
  \caption[]{MDS/MAGIC survey fields}
  \label{tab1}
  \begin{tabular}{c|cc}\hline
  Field & R.A.(J2000.0) & Dec.(J2000.0)\cr
  \hline\hline
	uqk00	& 07 42 44.0 & +65 06 08.0\cr
	usp00   & 08 54 16.0 & +20 03 41.0\cr
	uzp00   & 11 50 29.0 & +28 48 27.0\cr
	uzk03   & 12 10 34.0 & +39 28 53.0\cr
	uzx01   & 12 30 54.0 & +12 19 05.0\cr
 & & \cr
	uz 00   & 13 00 23.0 & +28 20 13.0\cr
	u26x1	& 14 15 20.0 & +52 03 01.0\cr
	u26x2	& 14 15 14.0 & +52 01 50.0\cr
	u26x3	& 14 15 07.0 & +52 00 40.0\cr
	u26x9	& 14 17 23.0 & +52 25 13.0\cr
 & & \cr
	u26xa	& 14 17 17.0 & +52 24 03.0\cr
	u26xb   & 14 17 10.0 & +52 22 53.0\cr
	u26xc	& 14 17 04.0 & +52 21 43.0\cr
	u26xd	& 14 16 57.0 & +52 20 33.0\cr
	u26xe	& 14 16 51.0 & +52 19 23.0\cr
 & & \cr
	u26xf	& 14 16 44.0 & +52 18 13.0\cr
	u26xg	& 14 16 38.0 & +52 17 03.0\cr
	u26xh	& 14 16 31.0 & +52 15 53.0\cr
	u26xi	& 14 16 25.0 & +52 14 43.0\cr
	u26xj	& 14 16 18.0 & +52 13 32.0\cr
 & & \cr
	u26xk   & 14 16 12.0 & +52 12 22.0\cr
	u26xl   & 14 16 05.0 & +52 11 12.0\cr
	u26xm   & 14 15 59.0 & +52 10 02.0\cr
	u26xn   & 14 15 52.0 & +52 08 52.0\cr
	u26xo   & 14 15 46.0 & +52 07 42.0\cr
 & & \cr
	u26xp   & 14 15 39.0 & +52 06 31.0\cr
	u26xq	& 14 15 33.0 & +52 05 21.0\cr
	u26xr	& 14 15 27.0 & +52 04 11.0\cr
	ux400   & 15 19 41.0 & +23 52 05.0\cr
  \hline\hline
  \end{tabular}
  \end{center}
\end{table}


\section{Introduction}

With the advent of the Hubble Space Telescope, observational cosmology
has been capable of extending the study of galaxy formation and evolution
to epochs corresponding to less than one third of the age of the Universe. 
The extremely narrow PSF produced by HST/WFPC2 images (around 0.1 arcsec) 
allow morphological classifications of galaxies at medium to high redshifts. 
The Medium Deep Survey key project \cite{gri94} has been classifying 
field galaxies from hundreds of WFPC2 images, providing photometry in
two bands (F606W and F814W, which roughly correspond to standard $V$ and $I$, 
respectively) as well as morphology down to $I\sim 22.0$. 
These data allow the study of number counts selected by type 
in $I$ and $V$ bands (Glazebrook et al. 1994, Driver et al. 1995, 
Abraham et al. 1996). The results indicate that counts in elliptical 
and spiral galaxies match closely the predictions of non-evolving models,
whereas the population of irregular/peculiar galaxies presents an 
excess over no-evolution estimations. The median redshift for 
spectroscopically confirmed galaxies is $z=0.5$. 

Inside knowledge of formation and evolution processes
conforming the observed field galaxy population mixture requires additional
data. Spectroscopic studies will provide intrinsic luminosities as well as
information about absorption and emission lines. Besides, optical--infrared 
colours are sensitive to the stellar population content. Optical emission 
comes mostly from young OB stars while the old stellar population 
dominates the infrared luminosity. In a recent paper, Glazebrook et al. 
\shortcite{gla98} study the characteristics of the faint MDS population 
through optical spectroscopy and near-infrared photometry. An excess of 
blue ellipticals is found in $I-K$ distributions compared to pure 
luminosity evolution models. The main purpose of this work is to study 
the characteristics of normal galaxies in MDS fields through 
optical-infrared colours.

This paper is organized as follows: A detailed account of the observations
and data reduction processes is presented in Section~2. We present
the number counts in $K^\prime$ band, compare them with
previous observations and discuss the morphologies 
contributing at different magnitudes, in Section~3. Sections~4 and 5 are 
dedicated to optical--infrared colour relations, the latter focusing
on the blue spheroid population found. A search for Extremely Red Objects
is shown next. Finally, a discussion of our results is presented in 
Section~7.

\section{Observations \& Data Reduction}
The targets for our $K^\prime$--band survey were chosen from the 
Medium Deep Survey catalogue that can be accessed at the
HST/MDS Archive (http://archive.stsci.edu/mds/mds.cgi). 
Our search was based on optimal positioning in the sky for our observing 
time and place, taking extra care not to bias the catalogue towards 
regions nearby quasars, active galaxies or other peculiar objects.
We found a set of images taken under HST program 
GTO5090 (E.Groth as Principal Investigator) out of which 22 fields 
were selected. $K^\prime$-band images of the MDS fields were 
obtained at the 2.2m MPIA telescope on Calar Alto, Spain during the 
nights of 1997 February 18,20 and  May 21, with MAGIC, a NIR camera 
that uses a NICMOS3 $256^2$ Rockwell HgCdTe array, with a pixel scale
of 0.64 arcsec. We used the $K^\prime$ filter instead of standard 
$K$ to reduce the sky background, thereby 
reaching a fainter limiting magnitude. The difference between $K$
and $K^\prime$ stays below 0.01 mag if we consider the translation
formula of Wainscoat \& Cowie \shortcite{wain} and a colour 
$H-K$ of a typical galaxy, given by a synthetic model 
\cite{bruz98}. Hence, we will denote with $K$ our $K^\prime$ magnitudes 
hereafter. 

Five-second individual exposures 
were taken, adding up 12 of them into a single 1 minute file.
Between frames we nodded the telescope in steps of 10 arcseconds,
following a path that closed in 9 steps. Each field has three such
sets totaling 27 minutes of integration time. Several
UKIRT faint photometric standard stars were also imaged throughout each
night. The fields were flatfielded using the median 
of the set of images for each field excluding that of the frame
being processed. The complete sample comprises 29 fields (table~1)
which covers a 116.8 arcmin$^2$ region in the $K$-band, with a 
89.4 arcmin$^2$ overlap with the HST/WFPC2 images. A translation
from the HST/WFPC2 filters to standard $V$ and $I$ photometry is done
using the transformation formulae from Holtzman et al. \shortcite{hol95}
and approximating $V-I$ to F606W$-$F814W. The additional correction terms
are less than 0.01 mag.

\begin{figure}
 \epsfxsize=3in
 \epsffile{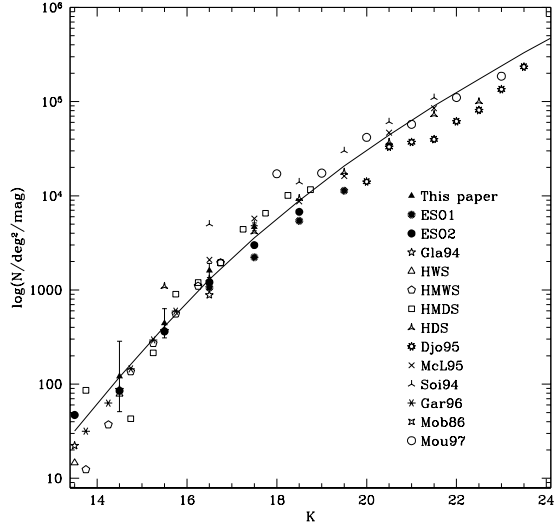}
 \caption{$K$-band number counts compared with previous surveys. The solid
 line corresponds to a no-evolution model. The solid triangles represents our
 results for objects detected {\sl both} in $I$ ($<22$) and $K$ ($<18$).
 Poisson error bars are shown.}
 \label{f1}
\end{figure}

The object detection process was done using SExtractor \cite{bert96}
using a $2.5\sigma /\sqrt{N}$ detection threshold.
N is the number of connected pixels within one seeing disk. Hence, 
the global threshold is $2.5\sigma$. Even though this value  might be rather
low for individual band detection, we are only considering those objects
that appear {\it both} in the $K$ and MDS-$I$ or $V$ images. 
We compared the aperture and isophote-fitting photometry from SExtractor and
IRAF's QPHOT task using 4, 5 and 6 arcsecond disks and found the maximum
uncertainty to lie around $0.2^m$ down to a limiting magnitude of $K=19$.
Eventually, we decided to use a 6 arcsec aperture photometry, which
guarantees all of the flux from each object is measured. Moreover, 
this is the same aperture used by Glazebrook et al. \shortcite{gla98},
which allows a better comparison. 

In order to match the WFPC2 and MAGIC images, we retrieved individual
HST frames for each field from the STScI archive (with exposure times around
1,000 seconds) and compared them to our NIR images.
We took several objects in each field to find the pixel-to-coordinate
mapping. Many of the images used in the survey overlap, and so we
checked for objects appearing twice.
Table~2 shows the number of objects detected as well
as the ratio of spheroids and disks in the sample (Bulge to Disk ratio
$>0.75$ and $<0.25$ respectively). As expected, the bias in imposing
detection in all bands: $V$,$I$ and $K$ implies a higher proportion 
of disks, compared to only imposing $I$ and $K$ detection.

The completeness magnitude was estimated in two different ways: A quick
method involves searching for the peak in the histogram of objects
appearing both in $K$ and $I$ bands to make sure we were rejecting spurious
detections. This method is only possible as long as we are 
working with images that are much deeper in $I$ than in $K$, so that no
severe bias is included by enforcing $I$ band detection.
Besides, the peak of the histogram will only show the completeness 
level as long as the Luminosity Function is monotonically increasing
in the detection range (which is our case; see, for instance, fig.~1).
The second method consists of simulating a random field of
pointlike galaxies with the Luminosity Function obtained from the
literature (cf.~fig~1) and with the same sky noise and pixel size. 
We could have considered a sample of spheroids and disks in the simulated
sample as the threshold of detection deals with surface brightness; however,
most of the galaxies from the MDS catalogue had half-light radii
well below 1.5 arcsec (see, for instance, table~3) which was our
resolution taking into account seeing conditions and pixel size. 
Both methods yielded a completeness magnitude of $K=18.0$.

\begin{figure}
 \epsfxsize=3in
 \epsffile{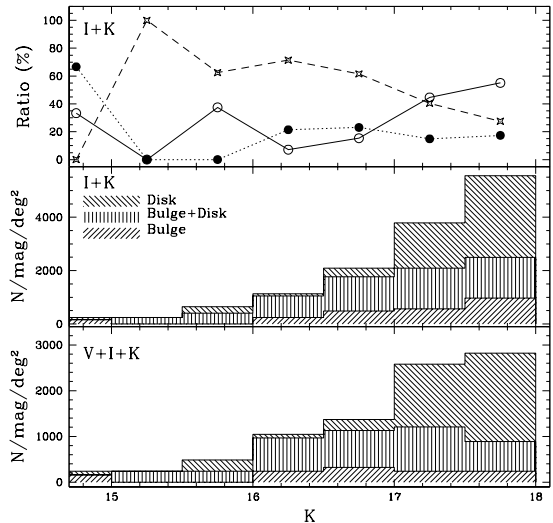}
 \caption{Evolution of morphology types as a function of $K$-band apparent
	magnitude (i.e. approximately as a function of redshift). 
	The top panel gives the ratio of each type to the total: 
	Solid circles: Bulges, Stars: Bulge+Disk, Hollow circles: Disks.}
 \label{f2}
\end{figure}

\begin{table}
  \begin{center}
  \caption[]{Survey Properties}
  \label{tab2}
  \begin{tabular}{c|cc|cc}\hline
  Band & Comp.Mag. & Objects & Spheroid(\%) & Disk(\%)\cr
  \hline\hline
  F606W ($V$)	  & $V<21.8$ & 276 &  9.8 & 62.7\cr
  F814W ($I$) 	  & $I<21.0$ & 377 & 15.1 & 53.8\cr
  $I$ + $K$	  & $K<18.0$ & 170 & 18.0 & 40.0\cr
  $V$ + $I$ + $K$ & $K<18.0$ & 109 & 14.0 & 45.0\cr
  \hline\hline
  \end{tabular}
  \end{center}
\end{table}


\section{NIR number counts and morphology}

We construct the number counts in the NIR with the objects
detected by SExtractor in the MAGIC images which have counterparts
in MDS $I$ band frames. Even though this might appear as a severe
bias on the sample, it is not, as can be seen in figure~1, where
our $I+K$ sample is shown along with data from the literature. 
A strong bias would appear 
as a paucity of objects and thus a disagreement with previous 
$K$ band surveys. The only population that might drop out in this process
would consist of blue irregular galaxies, which contribute negligibly
to the Luminosity function in our magnitude range ($K\le 18$).

The counts follow a $dlogN/dm$ relation
with a slope of $\sim 0.4$ in the range $16\le K \le 18$.
Differential number counts versus $K$ magnitude of our
sample are presented in figure~1 (solid triangles), along with Poissonian
error bars.
Just for comparison, a no-evolution model (solid line) based on    
the luminosity function given by Efstathiou et al. \shortcite{efs88} 
is also presented.
We assume a cosmology with $\Omega_0=0$ and $H_0=50$ km/sec/Mpc 
and a galaxy  mixture (including SFRs and IMFs) as 
prescribed by Pozzetti, Bruzual and Zamorani \shortcite{poz}. The
$k$-corrections and colours are calculated for different types 
of galaxies using the spectral synthesis code from Bruzual and Charlot 
\shortcite{bruz98}. The model has been normalized to the observations from the 
literature at $18<K<18.5$ (the present sample not included). 
The no-evolution model is compatible with the observations down to our 
completeness magnitude. Moreover, the NIR number
counts down to $K\sim 20$ cannot discriminate between no-evolution 
and passive luminosity evolution models \cite{poz}. 
On the other hand, the slope of the $K$ band number counts seems to 
flatten for the faintest magnitudes (Djorgovski et al. 1995,
Moustakas et al. 1997).

\begin{figure}
 \epsfxsize=3in
 \epsffile{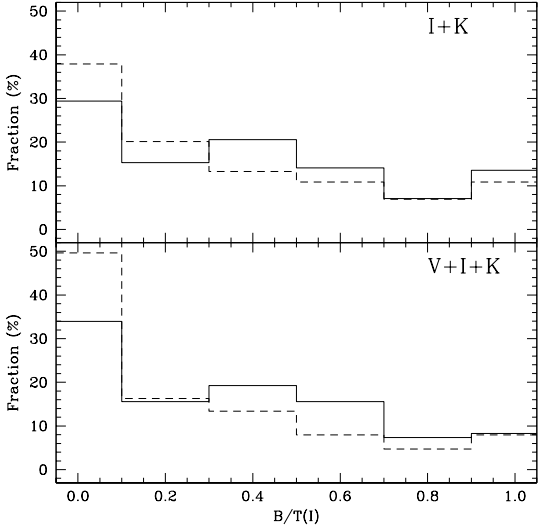}
 \caption{Bulge-to-Total fractions in $I+K$ and $V+I+K$ selected samples
 (solid lines). The dashed lines represent the histogram without the
 constraint of $K$ band detection.}
 \label{f3}
\end{figure}

Unlike many previous $K$ band surveys, now we have access to the morphology
of the objects from the $I$ and $V$ band images classified by the
MDS group. Figures~2 and 3 serve as a test of the contribution from  each
morphology to the counts in different bands.  We plotted in figure~3 the 
fraction of the total number of galaxies detected in $I$ and $K$
(top) and in $V$, $I$ and $K$ (bottom) versus the bulge to total luminosity 
ratio measured in $I$. As in a typical field sample, the
proportion of disk galaxies is higher than that of bulges.
The percentage of the latter ($B/T>0.75$) is smaller in the sample 
selected in $V$,$I$ and $K$ than in the one 
selected in $I$ and $K$ alone. This result follows the ratios observed in
the MDS objects in the two filters $I$ and $V$ (see table~2).
From figure~2 (top), one can notice that the ratio of bulges in
field galaxies decreases as we go to fainter magnitudes, down to 
$\sim 1/10$ at $K = 18.0$, whereas the disk population comprises
roughly 60\% of the sample at the faint end.
The contribution of the different populations to the differential
number counts at different $K$ magnitudes is presented in 
figure~2 (middle and bottom panels). 
One can see in both figures the trend indicated above:
faint magnitude counts are clearly dominated by disks. 


\section{Colour--Magnitude relations}

Figures~4a and 4b show the colour--magnitude relation for our sample in
$V-K$ and $I-K$. Having the morphology of each object, we can go further
and predict the colours that galaxies with a given morphology should have.
The shaded regions in these figures show the range of colours expected
for galaxies with redshift $0<z<1$, according to the models of 
Bruzual \& Charlot \shortcite{bruz98} for two metallicities:
one higher above solar ($2.5Z_\odot$) and another lower than solar
($Z_\odot /5$). 
All of them have a Scalo IMF defined for $0.1 M_\odot < M < 100 M_\odot$
and an exponential Star Formation Rate with a characteristic time $\tau$.
The morphology determines the value of $\tau$: 1 Gyr for bulges, 
5 Gyr for bulge + disk and $\infty$ (i.e. constant SFR) for disks. 
The formation redshift for all three models is $z_F=5$. Given the
degeneracy which exists between age and metallicity 
(e.g. Worthey 1994), we can account
for a range of colours either by changing the formation redshift or by
using different metallicities. In the figures we used the latter: we can
see a lower metallicity shifts the colour blueward in the same way a
lower formation redshift would. 

\begin{figure}
 \epsfxsize=3in
 \epsffile{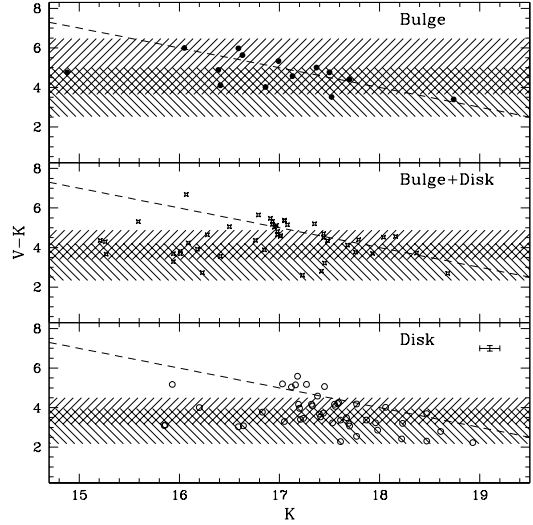}
 \caption{{\bf a.} $V-K$(a) and $I-K$(b) 
	colour--magnitude relation. The shaded area spans the colour
	region predicted by a model from Bruzual \& Charlot \shortcite{bruz98}, 
	between
	redshifts $0<z<1$ for two metallicities: $2.5Z_\odot$ (redder
	shading) and $Z_\odot /5$ (bluer region). The dashed line shows 
	the completeness limit in $V$(a) or $I$(b). A conservative error
	bar (i.e. the typical error for objects at the completeness 
	magnitude) appears in the bottom panel. Details of the models 
	used appear in the paper.}
 \label{f4a}
\end{figure}

\setcounter{figure}{3}
\begin{figure}
 \epsfxsize=3in
 \epsffile{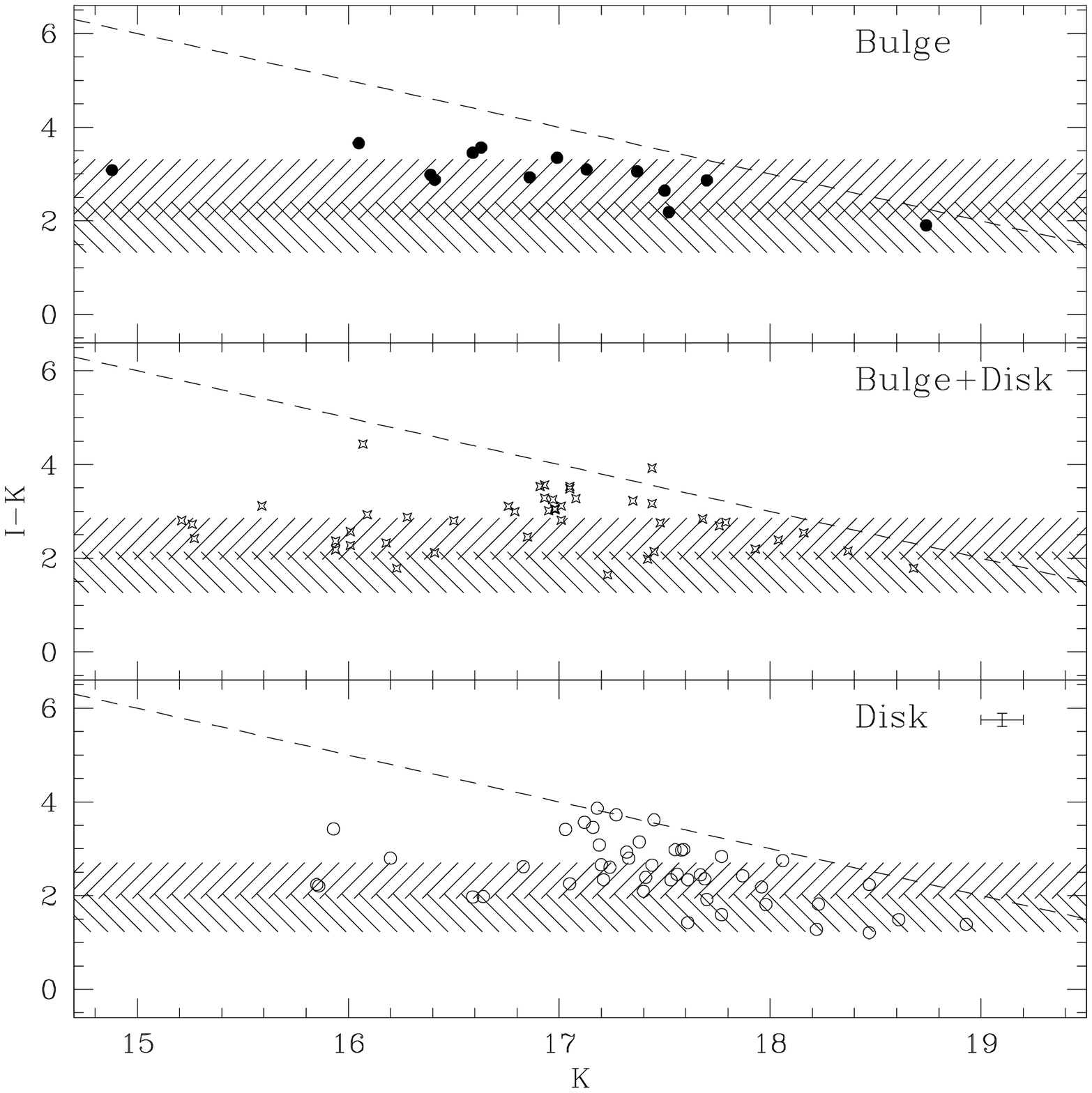}
 \caption{\bf b.}
 \label{f4b}
\end{figure}

Unfortunately, we do not know the redshifts of these galaxies and so 
we cannot compare the trajectory predicted by the models with the colours
of MDS galaxies. One is tempted, though, to use a $K-z$ relation from 
deep surveys \cite{cow96}; however the spread in this relation 
is too large to produce a meaningful result. A colour--colour plot 
avoids the estimation of redshift, and is shown in figure~5.
The dotted, solid and dashed lines correspond to
three different metallicities, namely: $Z/Z_\odot$=2.5,1.0 and 0.2, 
respectively. The solar metallicity curve also gives the redshift as
crosses, from $z=0$ (blue-blue i.e. lower left corner) up to $z=1$
in steps of $\Delta z=0.1$. There is good agreement with the 
colour--colour trend for all three morphologies, although a range of
metallicities (or of formation redshifts) is needed to account 
for quite a few of the galaxies. It is worth mentioning this issue
as either age or metallicity can account for this range of colours.
The few disks with a red $V-K$ and $I-K$ colours in figure~5 which
fall away from the model predictions should be considered with care
as they fall close to the detection limit in $V$ and $I$ and so
the photometric error in these bands could be large. The result
for spheroids is treated in the next section.

Finally, we left as an interesting exercise to compute 
the values estimated for the redshift using a purely photometric approach.
This method compares the broadband colours with templates associated with
the morphology of the galaxy, shifting the redshift until the predicted
colours are compatible with the observations. The more bands used, 
the better estimation one should have. Photometric redshift measurements
 can be thought of as a ``poor man's spectroscopy''.
Previous work on the subject has achieved remarkable accuracy:
Brunner et al. \shortcite{bru97} did a multiband analysis which resulted 
in a dispersion $\Delta z \sim 0.02$ up to $z=0.4$. A deeper estimation
of the uncertainties can be found in Hogg et al. \shortcite{hog98}.

In our case, we just have two colours ($V-K$ and $I-K$), which drove us
to use a simple method to estimate the redshift, namely: minimize
the squared difference between predicted and observed colours, using
for each morphological type the models described above with three
different metallicities. We imposed as a valid estimation that which
corresponded with a colour difference less than $\pm 0.5$ mag,
which reduced the sample to roughly 45 \%.
The histogram of the objects can be seen in figure~6, which has a 
median around $z\sim 0.2$. Even though the value for the complete
MDS sample is $0.5$, our lower redshift is reasonable as
the NIR survey done in the present work is only complete down to $K=18$.

\begin{figure}
 \epsfxsize=3in
 \epsffile{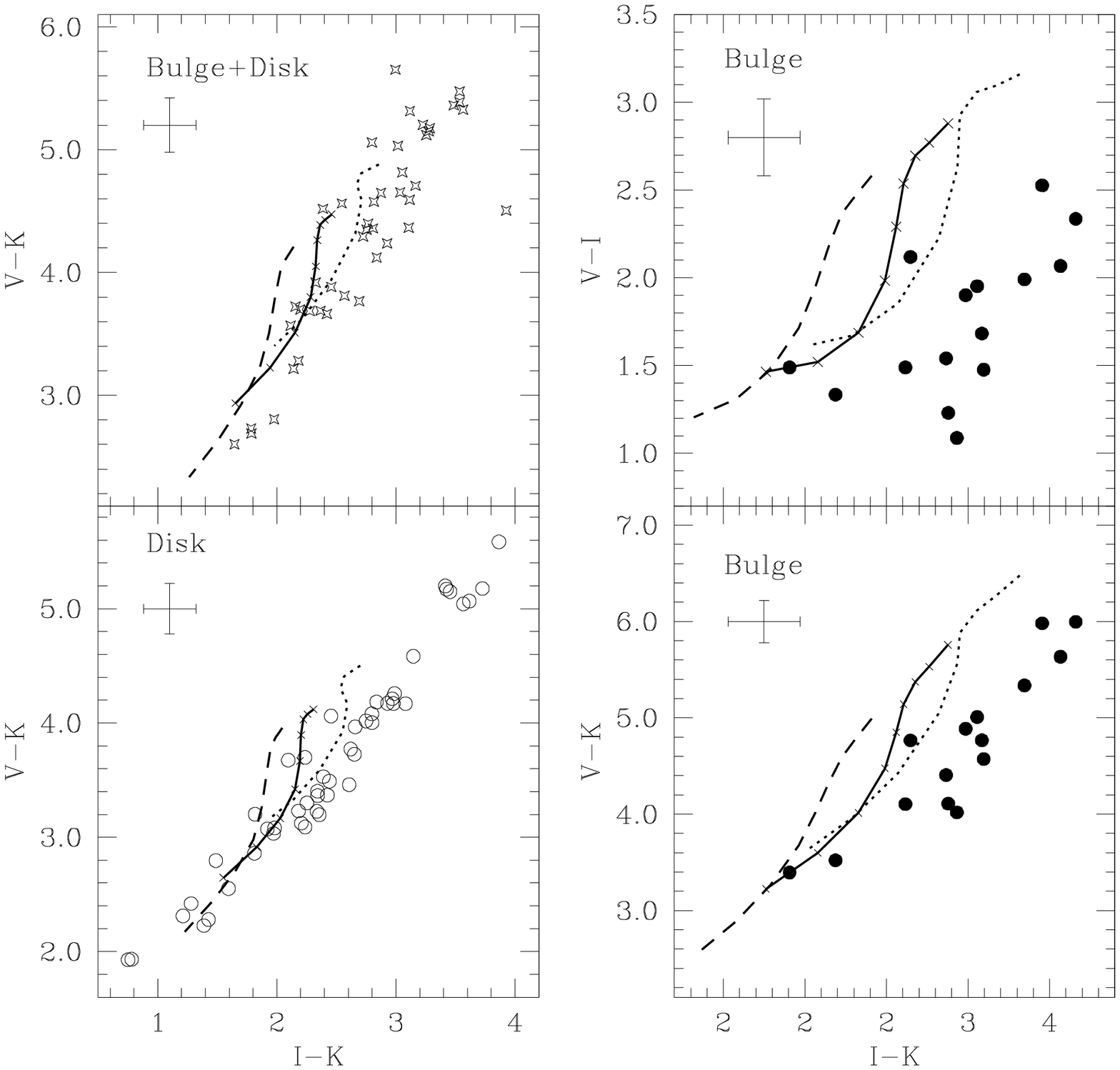}
 \caption{Colour---Colour plots. The lines correspond to model 
	predictions from Bruzual \& Charlot \shortcite{bruz98}
	for three different metallicities: 
	$2.5 Z_\odot$ (solid), $Z_\odot$ (dotted), and $Z_\odot /5$
	(dashed). A formation redshift $z_F = 5$ is assumed and
	each type only differ by their SFR timescale: $\tau = 1,5,\infty$
	Gyr for B, B+D and D, respectively. The typical error bar at the
	completeness level is also shown for each diagram.}
 \label{f5}
\end{figure}


\begin{table*}
  \begin{center}
  \begin{minipage}{70mm}
  \caption[]{Spheroids from MDS/MAGIC Survey}
  \label{tab3}
  \begin{tabular}{c|ccc|cc|c}\hline
  ID & F814W & F606W$-$F814W & F814W$-K$ & r$_{\rm HL}(arcsec)$ 
	& B/T & Field\cr
  \hline\hline
 E1 & 19.83 & 0.83 & 2.97 & 0.37 & 1.00 & uqk00\cr
 E2  & 19.75 & 1.80 & 3.70 & 0.52 & 0.82 & uzk03\cr
 E3  & 20.47 & 1.50 & 3.10 & 0.37 & 1.00 & uzk03\cr
 E4  & 19.42 & 1.46 & 3.03 & 0.49 & 0.76 & u26x1\cr
 E5  & 20.19 & 1.63 & 2.69 & 0.29 & 1.00 & u26x1\cr
 & & & & & & \cr
 E6  & 19.75 & 1.02 & 2.23 & 0.90 & 1.00 & u26x1\cr
 E7  & 20.24 & 1.59 & 3.61 & 0.32 & 0.79 & u26xq\cr
 E8  & 18.01 & 1.29 & 3.13 & 0.88 & 0.87 & u26xj\cr
 E9  & 20.27 & 1.13 & 3.14 & 0.29 & 1.00 & u26xj\cr
 E10  & 19.33 & 0.94 & 2.92 & 0.32 & 1.00 & u26xi\cr
 & & & & & & \cr
 E11  & 20.61 & 1.18 & 2.91 & 0.19 & 0.83 & u26xi\cr
 E12  & 20.08 & 1.95 & 3.49 & 0.53 & 1.00 & u26xg\cr
 E13  & 17.20 & 1.14 & 2.66 & 0.13 & 1.00 & u26xd\cr
 E14  & 20.38 & 1.53 & 3.39 & 0.34 & 0.85 & u26xd\cr
 E15  & 20.69 & 1.14 & 1.95 & 0.58 & 1.00 & u26xc\cr
  \hline\hline
  \end{tabular}
  \end{minipage}
  \end{center}
\end{table*}

\section{Blue spheroids in the field}
The right panels of figure~5 plot the spheroids in two different
colour--colour diagrams. In both cases the observations fall blueward
of the passive evolution estimations from spectral evolution models.
This is proof of the existence of ongoing star formation. Besides, 
the mismatch is bigger in $V-K$, than in $I-K$, possibly a sign that 
there is more flux into $V$ --- i.e. restframe $U$ and $B$ --- accounted for 
by young stars. Although the environment is totally different, 
this blueness is in agreement with the colours of some cluster 
ellipticals  (e.g. Couch \& Sharples 1987)
for which a recent episode of star formation must be considered in order 
to explain it \cite{cha94}. Confirmation of the existence of blue ellipticals 
in MDS fields appears in spectroscopic observations using the Keck 
telescope (Forbes et al. 1996, Koo et al. 1996).
We inspected visually all the spheroids in our sample and found only one 
possible misclassified object which appears as a late-type galaxy. 
The list of all 15 spheroids with their properties appear in table~3. 
These blue early-type galaxies should be further studied as they might 
throw light on the galaxy formation process both in the field
and in clusters.


\section{Searching for EROs}
A search for Extremely Red Objects (EROs) was also undertaken on this
sample. First of all we checked that none of the objects found 
{\sl both} in the $K$ band images and the HST/MDS catalog from the 
$I$ band frames had a colour characteristic of such sources 
($I-K > 4.5$). Next, we considered those objects from our $K$ band 
images alone, which did 
not appear in the MDS catalog and visually inspected both the
NIR and the HST/MDS images (the latter retrieved through the archive of
the STScI). About half of the candidates detected in $K$ fell in a region
which did not overlap with the HST $I$ or $V$ images. The other half had
counterparts which were either very bright stars or galaxies which
failed to adjust to the predetermined luminosity profiles. In either
case, these objects were brighter than $I\sim 21$ and so they
had $I-K<3$, i.e. bluer than the standard criterion for an Extremely
Red Object ($I-K>4.5$). Hence, all the objects detected in our $K$ 
band images --- at a completeness level of $K=18$ --- had counterparts 
in $I$ and had colours $I-K<4.5$. The sample extends over 89.4 arcmin$^2$
,which sets an upper limit for the number density of EROs at
$dn_{\rm EROs}/d\Omega < 0.011$ arcmin$^{-2}$ (for $K\le 18$).


\section{Discussion}

In this paper we have performed NIR photometry in 29 fields from the
Medium Deep Survey catalogue. The number counts obtained (figure~1) 
for objects detected {\sl both} in $I$ and $K$ band frames agree well
with previous $K$ band surveys. This agreement is possible since the
$I$ band images are deeper. The morphology obtained with high resolution
images from HST/WFPC2 allows a segregated study of colour-magnitude
and colour-colour relations for three different types according to the
bulge-to-disk ratios. As expected, spheroids appear redder in $V-K$ and
$I-K$ than disks and there is a fairly good agreement with spectral evolution
models \cite{bruz98} using a formation redshift $z_F=5$ 
and taking into account a sensible range in metallicities 
($0.2<Z/Z_\odot < 2.5$), which is equivalent to assuming a range 
in formation redshift, keeping a fixed metallicity. 
A simple photometric redshift estimation was done, giving 
a median redshift $z\sim 0.2$ which is reasonable given the 
completeness level of the $K$ band images. A further analysis
along these lines would require fluxes in additional bands.

In agreement with previous observations, we found spheroids
appear bluer than the model prediction for passive evolution, 
clearly showing the existence of ongoing star formation, similarly
to the blueness detected in some cluster ellipticals. Further study 
of this epoch of possibly strong star formation in the life of
early-type galaxies will be of great importance in the understanding
of galaxy formation.

\begin{figure}
 \epsfxsize=3in
 \epsffile{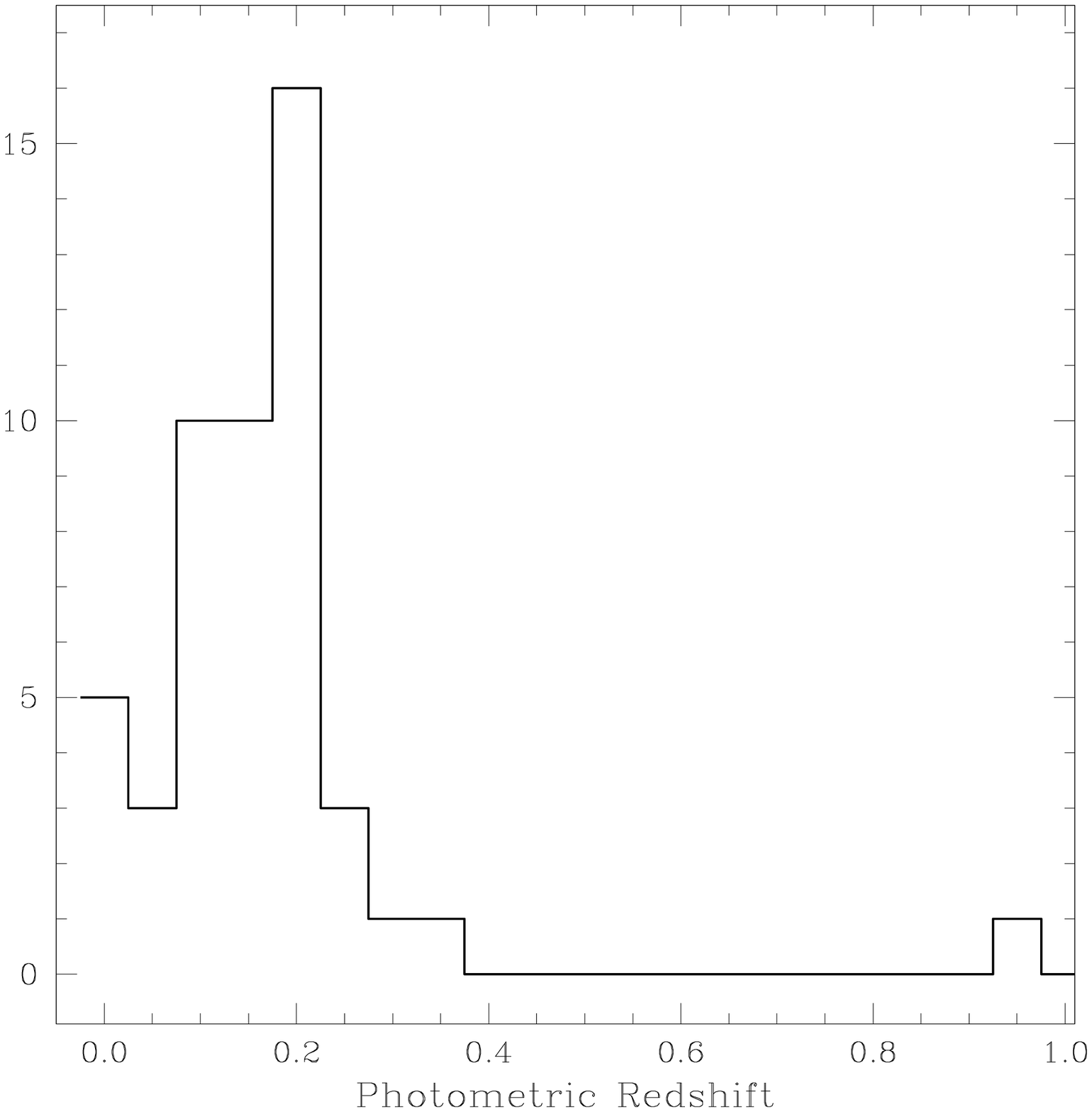}
 \caption[Figure 6.]
	{Histogram of the photometric redshift estimated for objects
	detected in $V$, $I$ and $K$ using the prediction for $V-K$ 
	and $I-K$ from the models of Bruzual \& Charlot \shortcite{bruz98}.
	The median redshift is $z_{\rm PHOT} \sim 0.2$}
 \label{f6}
\end{figure}

In order to compare our results with previous NIR surveys in MDS
fields \cite{gla98}, figure~7 shows the histogram of
objects detected in $I$ and $K$ along with the data from Glazebrook
et al. cutting their list at our completeness level. There is very good
agreement between both surveys, normalising to the same area.
A rank correlation test applied to 
both histograms yield a 100\% confidence level. It is also interesting
to notice that the extra objects in Glazebrook's sample as they
go deeper in $K$ appear blueward of the peak, signalling the existence
of galaxies with a stronger star formation rate, as expected for higher
redshift galaxies.

Finally, a search for EROs was undertaken on the overlapping $89.4$
arcmin$^2$ region which had photometry in $V$,$I$, and $K$. The whole
sample down to $K=18$ had counterparts in the F814W frames, all with
unconspicuous $I-K<4.5$ colours. Hence, no ERO was found, 
yielding an upper limit for the number density of roughly
 $dn_{\rm EROs}/d\Omega < 0.011$ arcmin$^{-2}$ ($K<18.0$), 
which agrees with previous estimates \cite{hu94}. 

\begin{figure}
 \epsfxsize=3in
 \epsffile{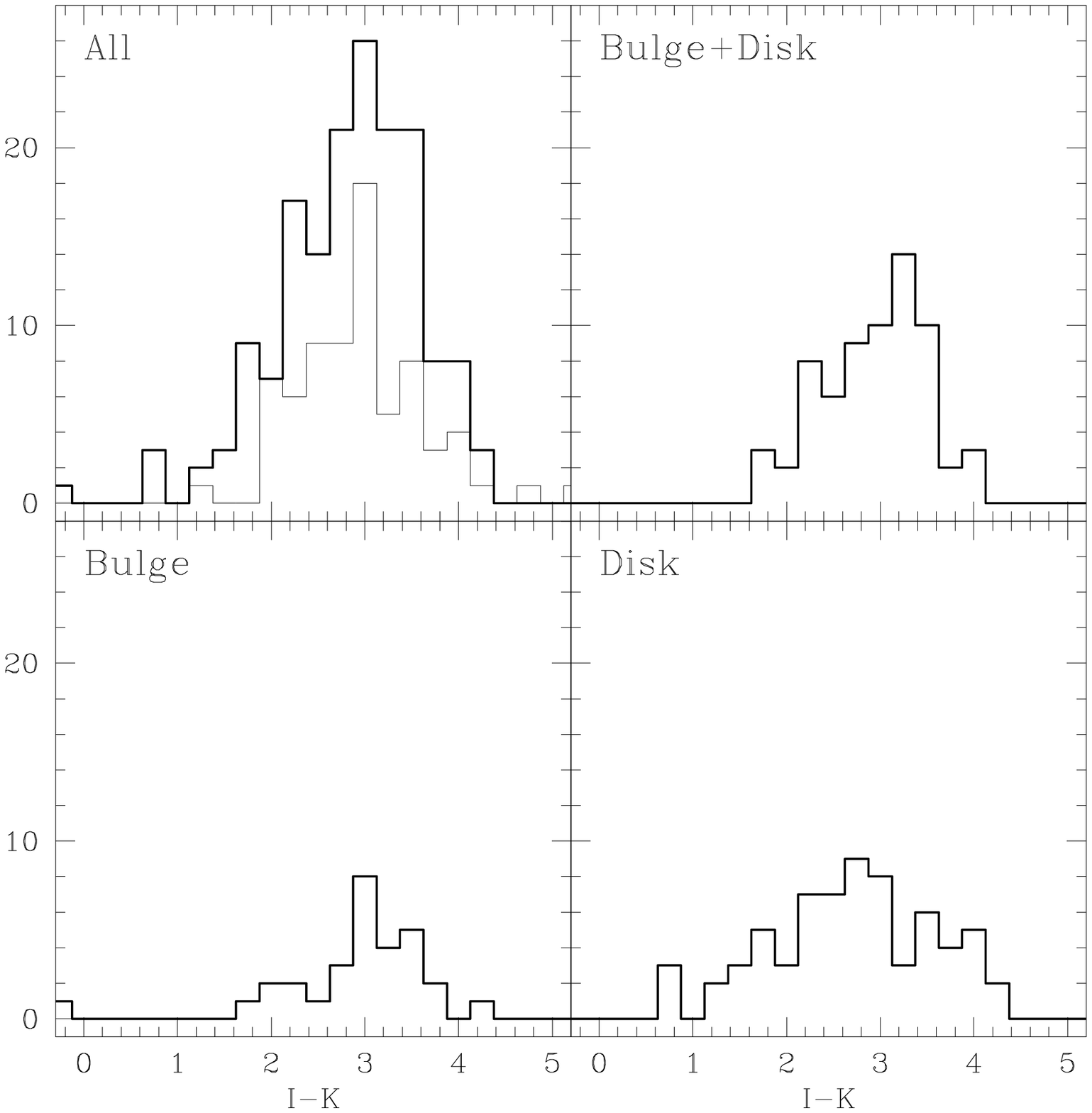}
 \caption[Figure 7.]
	{$I-K$ colour histogram as a function of morphology. The thin line
	shows the histogram from Glazebrook et al. \shortcite{gla98},
	which is scaled down to fit the box limits.}
 \label{f7}
\end{figure}


\section*{Acknowledgments}

The authors are grateful to the anonymous referee for helpful suggestions,
S.~F.~S\'anchez for pointing out valuable hints, J.~M. Diego for the 
help provided during the last run of observations, and P.~Saracco for 
providing us with the number counts data presented in Figure~1. 
We also acknowledge helpful discussions with F.~Hammer and J.~Silk.
The 2.2m telescope is operated  by the Max Planck Institut f\"ur
Astronomie at the Centro Astron\'omico Hispano Alem\'an in Calar Alto
(Almer\'\i a, Spain).
The Medium Deep Survey catalogue is based on observations with 
the NASA/ESA 
Hubble Space Telescope, obtained at the Space Telescope Science 
Institute, 
which is operated by the Association of Universities for Research in 
Astronomy, Inc., under NASA contract NAS5-26555. The Medium-Deep Survey 
is funded by STScI grant GO2684. 
I.F. acknowledges a a Ph.D. scholarship from the 'Gobierno de Cantabria'.
N.B. is supported by a Basque Government fellowship.
The authors acknowledge financial support from the Spanish DGES under
contract PB95-0041.


\end{document}